\documentclass[dvipsnames,twocolumn]{aastex62}
\accepted{4/3/19, ApJ}
\shorttitle{Turning Gravitationally Lensed Supernovae into Cosmological Probes}
\shortauthors{Pierel \& Rodney}
\usepackage{multirow}
\PassOptionsToPackage{usenames,dvipsnames}{xcolor}
\newif{\ifchangetext}
\changetextfalse

\InputIfFileExists{\jobname-options}

\ifchangetext
  
  \newcommand{\changenote}[1]{\textcolor{blue}{ \bf #1}}x
\else
  
  \newcommand{\changenote}[1]{}
\fi

\hypersetup{
    colorlinks=True,
    citecolor=Black,
    linkcolor=black,
    filecolor=magenta,      
    urlcolor=cyan,
}

\def\xone{\ensuremath{x_{1}}}

\newcolumntype{L}{>{$}l<{$}} %
\newcolumntype{C}{>{$}c<{$}} %
\newcolumntype{R}{>{$}r<{$}} %
\begin{document}
\title{Turning Gravitationally Lensed Supernovae into Cosmological Probes}

\correspondingauthor{J.~D.~R.~Pierel}
\email{jr23@email.sc.edu}
\author{J.~D.~R.~Pierel}

\affil{Department of Physics and Astronomy, University of South Carolina, 712 Main St., Columbia, SC 29208, USA}
\author{S.~Rodney}

\affil{Department of Physics and Astronomy, University of South Carolina, 712 Main St., Columbia, SC 29208, USA}

\begin{abstract}
Recently, there have been two landmark discoveries of gravitationally lensed supernovae: the first multiply-imaged SN, ``Refsdal'', and the first Type Ia SN resolved into multiple images, SN iPTF16geu. Fitting the multiple light curves of such objects can deliver measurements of the lensing time delays, which are the difference in arrival times for the separate images. These measurements provide precise tests of lens models or constraints on the Hubble constant and other cosmological parameters that are independent of the local distance ladder. Over the next decade, accurate time delay measurements will be needed for the tens to hundreds of lensed SNe to be found by wide-field time-domain surveys such as LSST and WFIRST. We have developed an open source software package for simulations and time delay measurements of multiply-imaged SNe, including an improved characterization of the uncertainty caused by microlensing. We describe simulations using the package that suggest a before-peak detection of the leading image enables a more accurate and precise time delay measurement (by $\sim1$ and $\sim2$ days, respectively), when compared to an after-peak detection. We also conclude that fitting the effects of microlensing without an accurate prior often leads to biases in the time delay measurement and over-fitting to the data, but that employing a Gaussian Process Regression (GPR) technique is sufficient for determining the uncertainty due to microlensing.
\end{abstract}

\

\section{Introduction}
\label{sub:intro}
The theory to enable the use of a gravitationally
lensed supernova (SN) resolved into multiple images as a cosmological
tool was developed in the seminal work of \citet{refsdal:1964}. As the light for each of the multiple images follows a different path through the expanding universe and through the lensing potential, the SN images appear delayed by hours (for galaxy-scale lenses) or years (for cluster-scale lenses). These time delays are sensitive to various cosmological parameters, enabling new measurements of the Hubble Constant $H_0$ and the dark energy equation of state \citep{Linder:2011,Treu:2016}. Time delay cosmography has been employed extremely successfully for decades using multiply-imaged quasars \citep[e.g.,][]{Vuissoz:2008,Suyu:2010,Tewes:2013b,Bonvin:2017,Birrer:2018}, but using gravitationally lensed SNe with multiple images (hereafter glSNe) can extend and enhance this method for several reasons.

glSNe have a strong luminosity peak and occur on short time-scales, enabling relatively simple and accurate time delay measurements \citep[e.g.][]{Woosley:2007, Sanders:2015}. In addition, the intrinsic luminosities of Type Ia and certain Type II SNe can be inferred independently of lensing \citep{Phillips:1993, 
  Hamuy:2002, Poznanski:2009, Kasen:2009}, which helps to determine the absolute lensing magnifications. This piece of information, not available for lensed quasars, provides an independent check on the lens model and the characterization of the line of sight, thus helping to break 
  the mass-sheet degeneracy \citep[e.g.,][]{ Holz:2001, Rodney:2015a}, which will otherwise introduce an additional uncertainty on $H_0$ \citep{Kolatt:1998,Xu:2016}.

The first multiply-imaged core-collapse \citep[SN Refsdal:][]{Kelly2015} and Type Ia \citep[SN iPTF16geu:][]{Goobar2016} SNe have been discovered in just the past few years. However, it is anticipated that the next generation of telescopes, particularly the Large Synoptic Survey Telescope (LSST) and the Wide-Field Infrared Survey Telescope (WFIRST), will provide hundreds to thousands of Type Ia and CC glSNe observations over the next decade \citep{Oguri:2010,Goldstein:2017}. With this enormous volume of observed glSNe it will become impractical to employ such individual treatments of time delay measurements seen in \citet{Rodney:2016} and \citet{Goobar2016}. Incidentally, the  time delay measurement techniques employed in these cases ignored the effects of ``microlensing'', whereby each SN image is separately varied by lensing effects from stars in the lens plane  \citep[e.g.,][]{Dobler2006,Bagherpour:2006,Foxley:2018}. 
This stellar-scale lensing operates at micro-arcsecond scales and can introduce uncertainty in the time delay measurement of $\sim4$\%, or $\sim2.4$ days for a time delay of 60 days.
\citep{Goldstein:2018}.

In this work we have developed an open source software package, written in Python, called \textit{Supernova Time Delays} (\textit{SNTD})\footnote{\href{https://ascl.net/1902.001}{SNTD}}. The SNTD package is capable of making accurate time delay measurements for glSNe of all types while including treatments of microlensing, and can immediately produce accurate simulations for wide-field time-domain surveys such as LSST and WFIRST. Section \ref{sec:sim} describes the simulation toolkit in SNTD including updated light curve models, the treatment of microlensing, and examples of SNTD-generated light curves using hypothetical lensed SN case studies. Section \ref{sec:delays} describes the time delay measurement capabilities of SNTD. This section uses the case-study lensing systems from section \ref{sec:sim} to assess the accuracy and precision of these measurements, and includes a comparison of time delay measurement accuracy for before-peak and after-peak SN detections.

\

\section{The SNTD Simulation Toolkit}
\label{sec:sim}
The simulation toolkit within SNTD allows a user to simulate realistic glSN observations with a variety of SN and lensing properties for various SN classifications. This capability can be used for design and optimization of surveys and follow-up campaigns. The software used inside SNTD for generating light curves and the ways in which it has been modified for use with glSNe are described in sections \ref{sub:sncosmo} and \ref{sub:sncosmo_ext}. Microlensing is described in section \ref{sub:micro}, both how it can be simulated and its inclusion in light curve simulations. In section \ref{sub:sim_cases} we use a representative lensed system to create simulated glSN observations, with microlensing included, to be analyzed for time delays in section \ref{sec:delays}.

\

\subsection{SNCosmo and Current Light Curve Templates}
\label{sub:sncosmo}

As discussed in section \ref{sub:intro}, SNe are an advantageous tool for time delay cosmography due to their light curve shape and time scales. Unlike the stochastic and heterogeneous light curves of AGNs, we have well-defined spectrophotometric time-series to describe a variety of SN types, with coverage from the ultraviolet (UV) to the near-infrared (near-IR). Within SNTD, these spectrophotometric time-series are used to fit observed glSN light curves, using the SNCosmo\footnote{\href{https://sncosmo.readthedocs.io/en/v1.6.x/}{SNCosmo Version 1.6}} python toolkit \citep{Barbary:2014}. SNCosmo has dozens of empirically defined time-series to describe the evolution of SNe Ia, Ib, Ic, II-P, II-L, and IIn, as well as the parametric SALT2 model for SNe Ia (see full table in the apendix). These SNCosmo models have coverage in the UV to optical wavelength range, and were extended by \citet{Pierel2018} to include more of the UV and near-IR. 

SNCosmo is a powerful tool for supernova cosmology, enabling users to fit existing models to observed data at any redshift while considering the effects dust extinction from any point along the line of sight. The toolkit is also capable of using any of the spectrophotometric time-series and fitting capabilities to produce realistic simulations of observed light curves. However, the toolkit is not designed for the analysis of lensed SNe, having no capabilities for handling macrolensing, microlensing, or glSNe. SNCosmo is also not equipped for the specific case of SN Refsdal, the first glSN discovered, which happened to be an unusual SN that cannot be fully described by existing SN classification templates \citep[e.g.][]{Arnett:1989,Kelly:2016}. 

\

\subsection{Extending SNCosmo}
\label{sub:sncosmo_ext}
As SNCosmo is already widely used by the SN research community, we have opted to extend the package to provide optimized glSN capabilities that work within SNTD, as opposed to creating an entirely new framework. The updates made are as follows:

\

\textit{Parametric light curve model}: while not uniquely useful to studies of glSNe, an important addition to the capabilities of SNCosmo is the inclusion of a parameterized model that operates within the framework of the existing package. This parameterized model allows a user to fit photometric data without knowledge of the SN classification or redshift, enabling time delay measurements in the cases where there is no spectroscopic follow-up or the data are not well fit with common light curve templates \citep[e.g.][]{Rodney:2016}. 

 A common practice for fitting photometric data with no underlying light curve template involves the use of splines or other flexible functions \citep[i.e.][]{Tewes2013}, but the simplicity of SN light curves enables the use of a parameterized model that has fewer free parameters. The parameterized light curve model defined by \citet{Bazin:2009} is implemented in SNCosmo for use in SNTD, which calculates flux as a function of time in such a way as to be general enough to fit any light curve shape. While useful for fitting light curves in certain situations, this parameterized model describes only the shape of a SN light curve; it does not have any information about the SN SED, so it cannot leverage information about the color of an observed SN. The \citet{Bazin:2009} model is defined according to:
 
 \begin{equation}
\label{eq:bazin}
F(t)=A\frac{e^{-(t-t_0)/\tau_{\rm{fall}}}}{1+e^{-(t-t_0)/\tau_{\rm{rise}}}}+B,
\end{equation}
where $\tau_{rise}$ and $\tau_{fall}$ characterize the relative rise and decline times of the light curve, while $A$ and $B$ are constants. By taking the derivative, it is apparent that the time of peak and peak flux are:
\begin{equation}
t_{\rm{max}}=t_0+\tau_{\rm{rise}}\ln\Big(\frac{\tau_{\rm{fall}}}{\tau_{\rm{rise}}}\Big)
\end{equation}
\begin{equation}
    f(t_{\rm{max}})=Ax^x(1-x)^{1-x}+B; \ x=\frac{\tau_{\rm{rise}}}{\tau_{\rm{fall}}}
\end{equation}
While this model is extremely flexible, it still mimics the smooth rise and fall of a general SN light curve, which helps to keep a light curve fit relatively insensitive to the effects of sharp microlensing features.

\

\textit{Microlensing:} we describe the phenomenon of microlensing, as well as SNTD's method for simulating it, in section \ref{sub:micro}. Once the simulation is complete however, the effect can be simply described by a time-evolved ``magnification curve''. This curve is added to a SN light curve as an SNCosmo ``propagation effect,'' in the same way as dust extinction is handled. These effects are simply models of how intervening structures affect a spectrum, as well as the associated light curve. The microlensing propagation effect causes time-dependent flux-scaling when applied to the simulated light curve points. Each image of a glSN will be along a different line of sight, subjecting it to unique microlensing effects. Therefore, during simulations, a separate microlensing effect is applied to each light curve as it is generated (See section \ref{sub:micro} and \ref{sub:sim_cases}).

\

\textit{Multiply-Imaged Supernovae:} knowing that observations of glSNe stem from a single light curve is extremely valuable in measuring time delays (see section \ref{sec:delays}). As SNCosmo has no structures for handling this phenomenon, SNTD contains new objects capable of maintaining maximum information about the SN and each of its images for use in time delay measurements. All components of the new objects operate within the current SNCosmo framework so that, for example, each SN image has a usable light curve within existing SNCosmo functions.

\

\subsection{Microlensing}
The phenomenon of microlensing occurs when light rays from the expanding photosphere of a SN pass within the lensing potential of a set of stars in the lens plane, which have Einstein radii on the order of micro-arcseconds \citep{Dobler2006}. 
This causes fluctuations in the light curve of $\sim0.2$ to $>0.5$ magnitudes on timescales of weeks to months, which can make it difficult to obtain an accurate time delay measurement without accounting for microlensing in certain lens configurations \citep{Dobler2006,Goldstein:2018,Foxley:2018,Bonvin:2019}. 

To simulate the effects of microlensing, we employ the \textit{microlens} inverse ray-tracing code \citep{Wambsganss1999}. The \textit{microlens} code defines random realizations of stars modeled as point-masses in the lens plane according to two parameters: $f_* \ \rm{and} \  q$. The parameter $f_*$ defines the relative fraction of convergence ($\kappa$) from discrete sources such as stars compared to smooth sources such as dark matter. The parameter $q$ is the stellar mass ratio ($m_{min}/m_{max}$) of the stellar mass function. The microlensing magnification distributions produced by \textit{microlens} require a further two parameters, namely the shear due to the Singular Isothermal Ellipsoid (SIE) and external potentials ($\gamma$), and the local convergence ($\kappa$). The convergence causes a focusing of light rays, leading to an isotropic magnification of a
source, while the shear introduces an anisotropy that distorts the shape of the magnified source \citep{Narayan:1996}.

\begin{figure}[ht!]
\centering
\includegraphics[ width=0.45\textwidth]{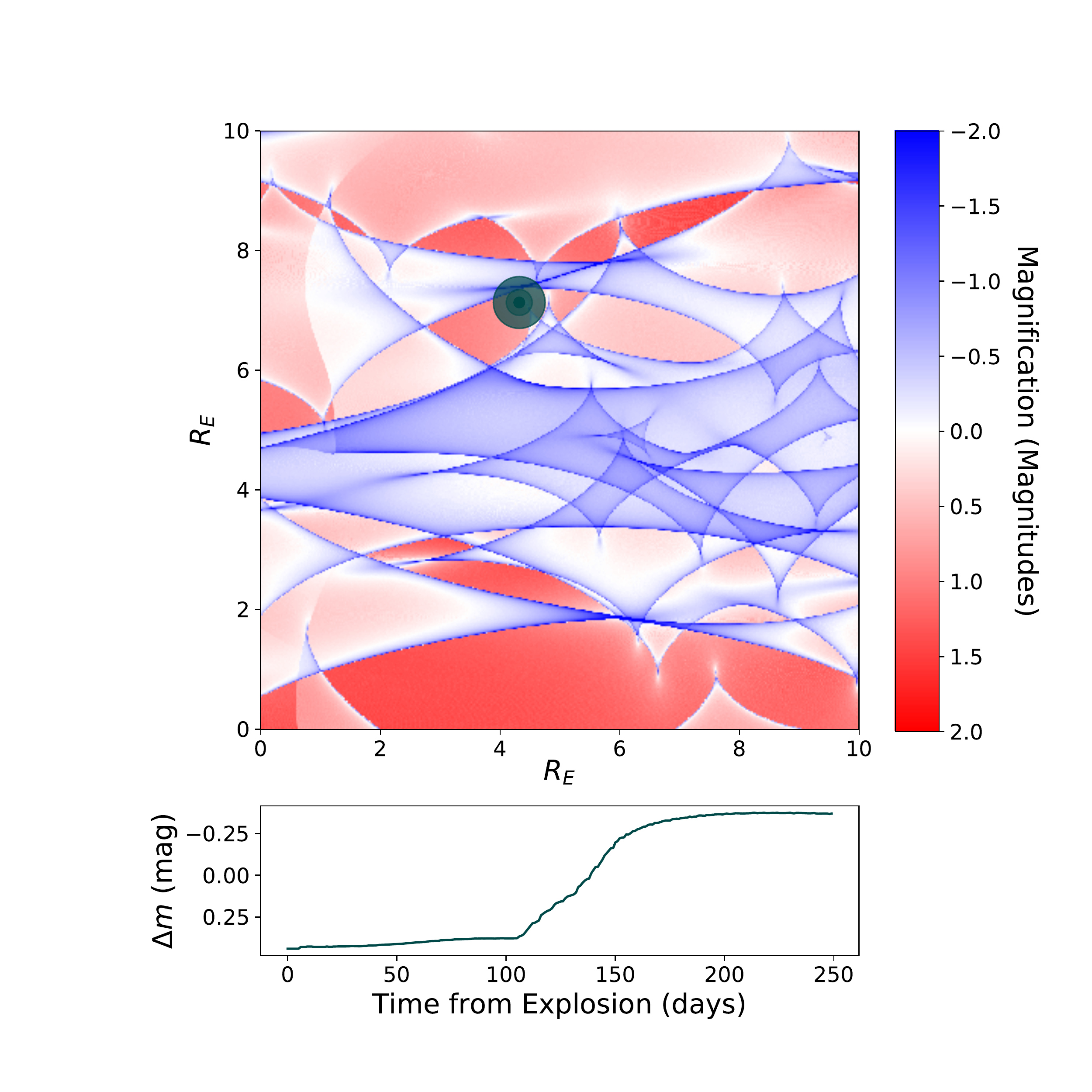}
\caption{\label{fig:microlensing} A microlensing microcaustic realization using SNTD, where caustics that increase magnification are blue and caustics that decrease magnification are in red. In this case, the lensing parameters described in section \ref{sub:micro} are set to $(\kappa,\gamma,f_*,q)=(0.31,0.61,0.5,0.2)$ and the photospheric velocity is set to $10^4\rm{~km~s^{-1}}$, slightly below the average peak photospheric velocity for SNe Ia \citep{Zheng:2018}. This photospheric velocity is nevertheless much larger than the peculiar velocities and proper motions in the lens and source planes, so we approximate the centroid of a supernova as constant with only the radius of the supernova changing with time \citep{Foxley:2018}.}
\end{figure}

Once the microcaustic field is generated by the \textit{microlens} code for a particular set of parameters, it must be convolved with a model for the expanding photosphere of a SN. By default, we create a model for an expanding SN photosphere as flat-disk achromatic Gaussian brightness distribution. This photospheric model is a simplifying assumption also made in previous microlensing analyses \citep[e.g.][]{Foxley:2018}, which other work has shown to be sufficiently accurate up to an ignored ``microlensing time delay'' on the order of 0.1 days \citep{Bonvin:2019}. The microcaustic field defines a grid of positive and negative magnifications in the source plane.  SNTD can produce a magnification curve---describing the effect of any given microlensing realization with time---by propagating the expanding photospheric model through the microcaustic field (Figure \ref{fig:microlensing}). As this default model is achromatic, the resulting magnification curve will be a function of only time. In principle SNTD is capable of replacing this simple model with a 2-D projected specific intensity profile, like that of \citet{Goldstein:2018}. This method would better represent a SN atmosphere, which emits light differently as a function of radius from the center \citep[e.g.][]{Kasen:2006} in such a way that a time delay effect can also be observed \citep[e.g.][]{Bonvin2018}. This substitution could be made by a user, or may perhaps be the default in a future iteration of SNTD. The SN image can be placed randomly at any position in the microcaustic field, so that one can explore the variety of microlensing effects that may arise for a given set of lensing parameters ($\kappa$, $\gamma$, $f_*$, $q$). The size of each microcaustic field projected onto the source plane is $10R_E$ on a side, where $R_E$ is the Einstein radius of a typical deflector of mass $M$ in the lens plane: 
\begin{equation}
    \label{eq:einstein}
    R_E=\sqrt{\frac{4GM}{c^2}\frac{D_{ls}D_s}{D_l}},
\end{equation}
where $D_s, \ D_l, \ \rm{and} \ D_{ls}$ define the angular diameter distances to the source, lens, and between the lens and source, respectively. 

\label{sub:micro}

\subsection{SNTD Simulation Examples}
\label{sub:sim_cases}
In order to demonstrate the simulation and time delay measurement capabilities of SNTD, we created 2,000 simulated glSNe using 
an imagined galaxy-scale strong-lensing system that is similar to the lenses described by \citet{Shu2018a}. It should be noted that a true SN detection in such a galaxy would require excellent follow-up observations from space and/or large ground-based observatories to obtain sufficient data quality for time delay cosmography. This mock lensing galaxy is placed at redshift ($z_{\rm{lens}}$) of 0.5, and the source galaxy is placed at a redshift ($z_{\rm{source}}$) of 1.4. The overall magnification factor ($\mu$) for the leading image of a SN located at the light peak of the lensed galaxy is set to 5, and the relative time delay ($\Delta t$) between the two images is set to 60 days. The measurement of time delays for these mock lensed SNe is described in section \ref{sec:delays}.

Each simulated SN has an absolute magnitude in the B bandpass defined by a Gaussian luminosity function (see appendix for table), a redshift equal to the host galaxy's $z_{source}$, and host galaxy extinction defined by the dust distributions used by \citet{Rodney:2014}. Magnification and time delay factors are applied to each SN image, defined by the lensing galaxy's $\mu$ and $\Delta t$ described above, respectively. 

A microcaustic field is randomly generated at a redshift equal to the lensing galaxy's $z_{lens}$ described above, following the methodology described in section \ref{sub:micro} using lensing parameters representative of the lenses presented by \citet{Shu2018a}. The resulting microlensing magnification curve is applied to each simulated light curve as an added propagation effect, causing positive and negative flux variations with time (Figure \ref{fig:microlensing}). 

To simulate realistic light curves the observational strategies presented by \citet{Shu2018a} are adopted, which describe a survey strategy that targets lensed galaxies with high star formation rates and prior knowledge of the lensing parameters. The photometric data are generated with observational parameters for the HST F125W and F160W filters and a fixed S/N curve based on the HST exposure time calculator. We simulate 1000 glSNe for which the SN is first detected before peak brightness, and another 1000 glSNe for which the first detection lands after peak brightness. With these two sets of simulations we can ask how much the precision of a SN time delay measurement is degraded when there is no detection before peak brightness. This will be an important consideration for the choice of cadence and depth in future surveys, and also for the allocation of resources for follow-up observations.

An example of an observations table used by SNCosmo to realize a set of light curves is given in the appendix. SNTD simulates the multiple images of a single SN by using the same absolute magnitude and host galaxy dust parameters discussed above for each light curve as they are inherent to the SN, then scaling the simulated flux measurements and shifting the time axis of the trailing image by the prescribed time delay and magnification parameters. Finally, the microlensing magnification curve described above and observational noise are applied to the simulated observations, causing further variations in the flux with time. 

\

\textit{Before Peak Detections:} We first generated light curves under the assumption that we were able to ``observe'' each leading SN image before its epoch of peak brightness. It is ideal to obtain this before-peak observation, so that the peak of the light curve is well-defined and more easily fit. A SN is simulated using the methods outlined above in this section, then a random epoch is chosen between the epoch at which the brightness of the SN reaches the detection threshold of the telescope, and the epoch of its peak brightness. This epoch defines the first observation of the leading image.

\begin{figure}[ht!]
\centering

\includegraphics[width=.45\textwidth]{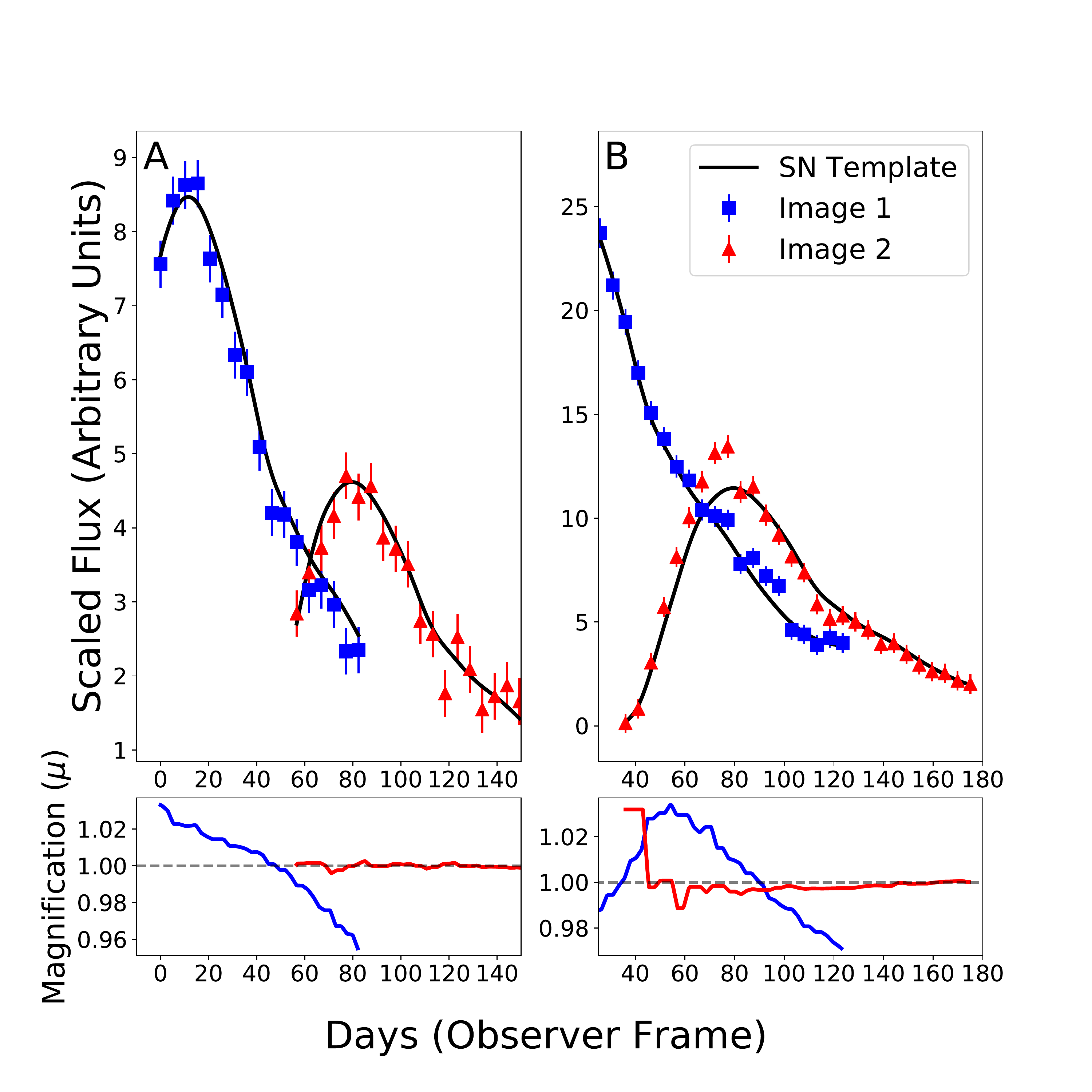}
\caption{\label{fig:case_sims} An example simulated glSN Ia from each of the case studies described in this section. In each panel, the solid black line is the underlying template used to generate the photometry before microlensing and observational noise are added, blue is used for the leading image, and red is used for the trailing image. The microlensing curves (see section \ref{sub:micro}) added to the light curves are shown in the lower panels. The left panel displays the two images of a SN for which the leading image was detected before peak (A), and the right panel displays the two images of a SN for which the leading image was detected after peak (B).
}%
\end{figure}

Once the SN is ``observed'', we assume a set of non-disruptive follow-up observations are made by the HST. This necessitates a 2-week gap between the first epoch and the second, and thereafter a 5-day cadence is used to ``observe'' the SN until the trailing image is no longer detectable (Figure \ref{fig:case_sims}A). This process is repeated to create a mock catalog of 1000 realistic glSN light curves.

\

\textit{After Peak Detections:} The process of generating a realistic catalog of 1000 glSN light curves is repeated under the assumption that we only detect each leading SN image after its epoch of peak brightness. It is more difficult to identify the peak of a light curve, and therefore the relative time delay, without any before-peak observations due to the lack of a primary inflection point.

The light curves for the glSNe detected after peak are generated by taking the first observation to be a random epoch between the epoch of peak brightness and the epoch at which the brightness of the SN falls below the detection threshold of the telescope. This detection epoch simultaneously defines the first observation of the trailing image, which is nearly always before peak in this case due to the relative time delay of 60 days. Then we again assume a set of non-disruptive follow-up observations are made by the HST. This results in a 2-week gap between the first epoch and the second, and thereafter a 5-day cadence is used to ``observe'' the SN until the trailing image is no longer detectable (Figure \ref{fig:case_sims}B).

\

\section{The SNTD Time Delay Measurement Toolkit}
\label{sec:delays}
We describe SNTD's treatment of microlensing in section \ref{sub:micro_fitting}, and then the various tools available to measure the time delays and relative magnifications of glSNe. The methods include parallel analysis (Section \ref{sub:delays_separate}), series analysis (Section \ref{sub:delays_combined}), and color curve analysis (Section \ref{sub:delays_colors}). Next, the simulated samples of glSNe created in section \ref{sub:sim_cases} are used to compare these three time delay measurement techniques.

\

\subsection{Accounting for the Effects of Microlensing}
\label{sub:micro_fitting}

\begin{figure}[ht!]
\centering
\includegraphics[width=.45\textwidth]{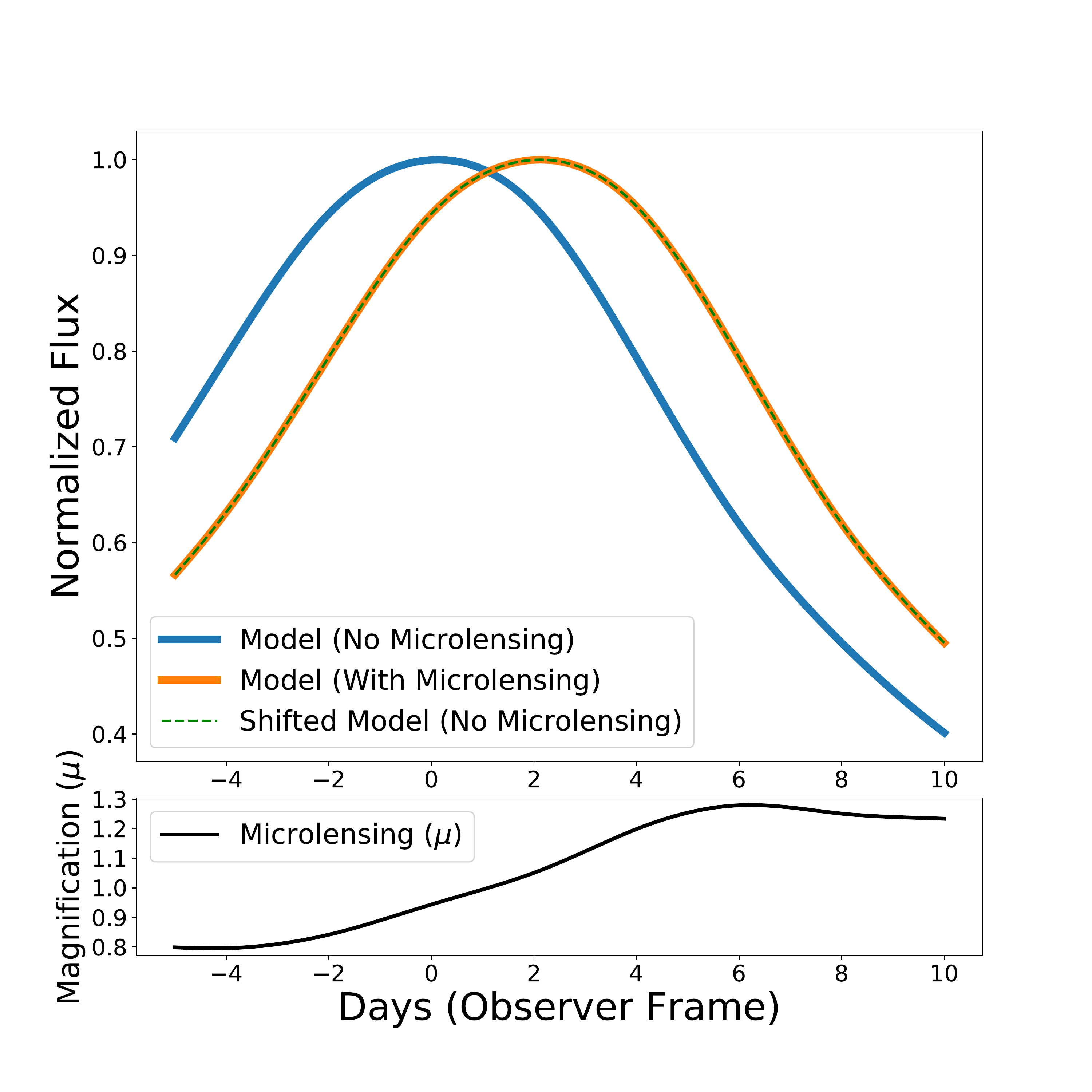}
\caption{\label{fig:micro_shift} A contrived case of microlensing where the shape of the light curve is untouched, but the time of peak is shifted by several days. First, a Type Ib supernova model produces a light curve with no microlensing added (blue). Next, a pathological but realistic microlensing curve (black) is applied to the non-lensed model. The resulting curve (orange) is identical to the non-lensed curve with the time of peak shifted (green). While this scenario is not likely, this idealistic case shows the extreme at which a microlensing curve can mimic a shift in the time of peak of a light curve, which would not be caught by any data-driven modeling process but contributes a systematic error to the time delay measurement.
}%
\end{figure}

Identifying and correcting for microlensing in a light curve is an extremely difficult problem, driven mainly by the stochasticity of the effect and the degeneracies present between the extrinsic and intrinsic model parameters. A pathological but plausible case of microlensing is capable of leaving the shape of a light curve relatively untouched, while shifting the peak in time (Figure \ref{fig:micro_shift}). The effect of microlensing from a modeling perspective would be classified as low in this case, as the model performs well in fitting the data, but there is an added systematic error unaccounted for in the measurement of the time of peak that will propagate through to an error in the time delay. 

\begin{figure}[ht!]
\centering
\includegraphics[width=.45\textwidth,trim={.75cm 0 .25cm 2.5cm},clip]{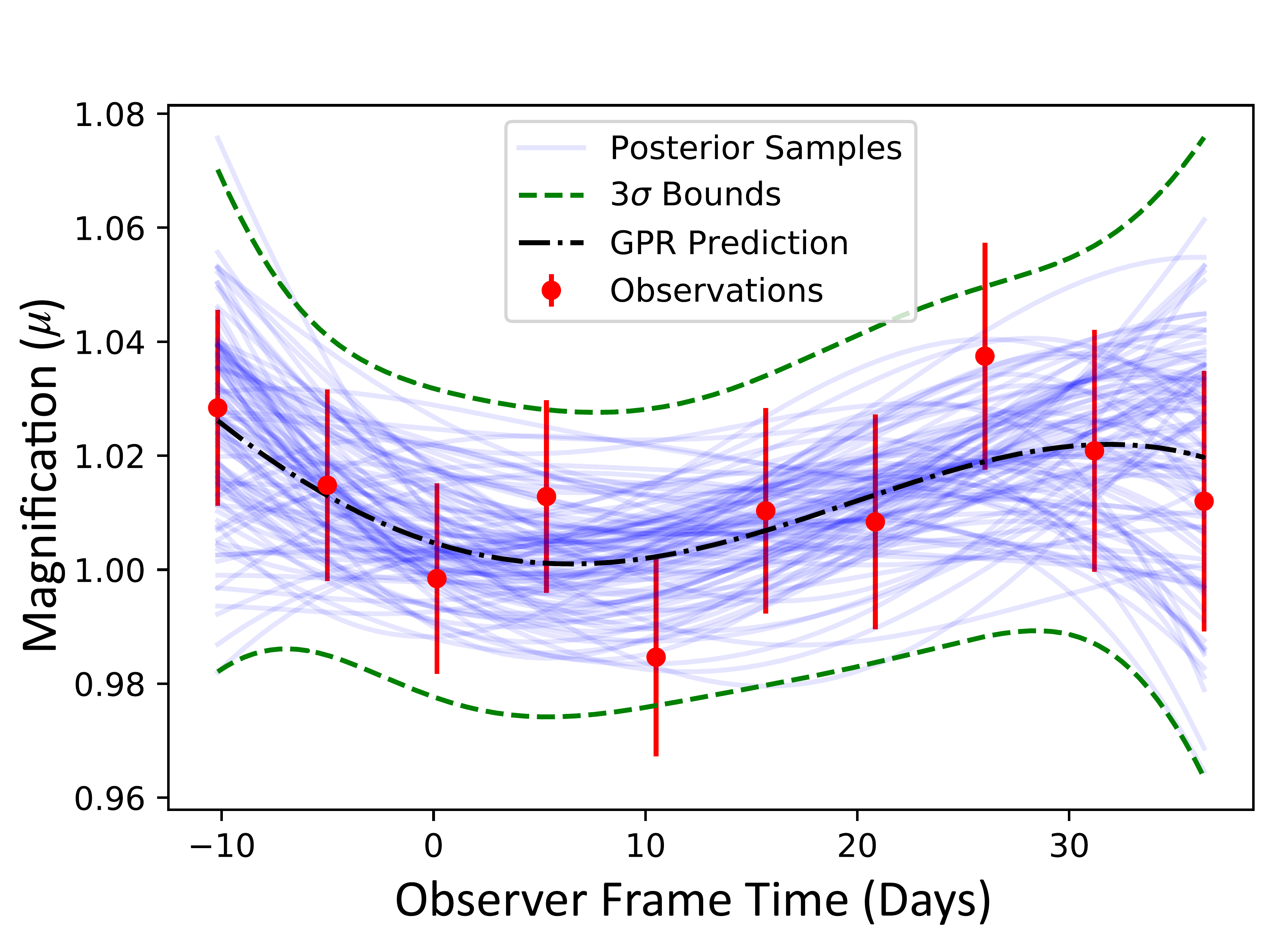}
\caption{\label{fig:micro_gpr} Red points with error bars are the residuals between an observed (simulated) SN light curve and a best-fit model. Light solid blue curves are the $\sim100$ samples from the GPR posterior used to represent a diverse range of possible microlensing scenarios. The dot-dashed black curve is the net GPR-predicted microlensing curve, and the dashed green curves represent the 3$\sigma$ bounds on that prediction. The data are refit with each of the light blue posterior samples included as the assumed microlensing curve to estimate the uncertainty due to microlensing (see section \ref{sub:micro_fitting}). }
\end{figure}

Therefore, we seek to better characterize the systematic error in the time delay measurement due to microlensing, which if unaccounted for would cause an underestimation of the time delay measurement uncertainty. When not in the microlensing regime described in figure \ref{fig:micro_shift}, fitting light curves for time delays without considering microlensing remains quite effective, due to the existence of templates for various SN classifications that reduce the flexibility of models. In these cases, the main deviations from the model should be due to microlensing effects, which are then visible in the residuals. However, simply fitting a curve to the residuals and attributing all of those deviations to microlensing is inadequate, because it will create a bias in your time delay measurement if your initial fit is poor, or may incorrectly attribute observational scatter to a microlensing effect \citep[e.g.][]{Tewes2013}. Therefore, we avoid directly fitting for a microlensing spline and instead use monte carlo simulations of realistic microlensing magnification curves to characterize the additional error on the time delay measurement:

\begin{enumerate}
    \item A best fit model for the intrinsic SN light curve is found by fitting observed light curve data with a model that has no microlensing included (see section \ref{sub:delays_separate}-\ref{sub:delays_colors}). This fit defines a measured time delay $\Delta t$.
    \item Trends in the residuals of the best fit model are identified using Gaussian Process Regression (GPR), which produces a posterior distribution of possible microlensing functions.
    \item A representative sample of N (typically 100) microlensing curves is then extracted from the GPR posterior (Figure \ref{fig:micro_gpr}). 
    \item  The photometric data is then refit once for each GPR posterior sample, with each microlensing curve applied to the best fit model as a microlensing propagation effect (section \ref{sub:micro}). The flux calculated by the model is affected by the microlensing propagation effect, and otherwise the same model parameters that are measured in step 1 are allowed to vary.
\end{enumerate}

\begin{figure}[ht!]
\centering
\includegraphics[width=.45\textwidth]{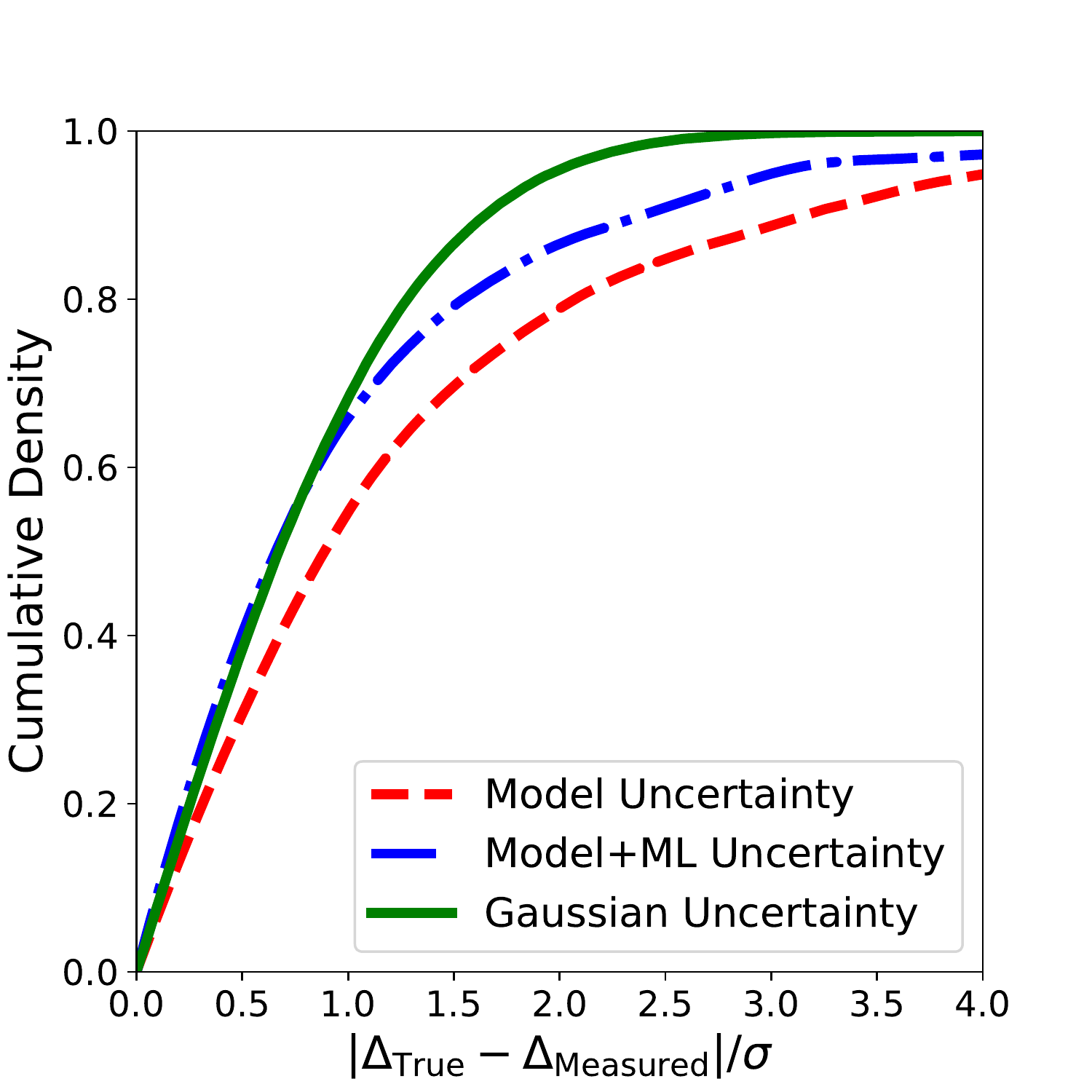}
\caption{\label{fig:microlensing_errors} Using the simulated SNe observed before-peak in section \ref{sub:sim_cases}, we seek to characterize the improvement on the time delay uncertainty reported by SNTD using the microlensing fitting technique described in section \ref{sub:micro_fitting}. The 1000 glSNe are fit with the technique described in section \ref{sub:delays_separate}, and then we restrict the SNe to those with a $\chi^2$ less than or equal to one corresponding to a P-value of 0.05. This subset represents SNe for which an observer would have a sufficiently accurate fit to believe their measurement result, but that have an unknown effect from microlensing. A value $n$ on the x-axis corresponds to the fraction of measurements that were $n\sigma$ from the actual time delay, where $\sigma$ is the total uncertainty. The three curves represent the cumulative densities for the following distributions: The red dashed curve includes only the model uncertainty, the blue dashed-dot curve includes the model uncertainty and the measured uncertainty due to microlensing, and the green solid curve corresponds to a perfectly Gaussian characterization of the uncertainty. } 
\end{figure}

The result of these steps is a time delay measurement error for each microlensing scenario, $\Delta t-\Delta t_i^\mu$, the combination of which defines a distribution of time delay errors due to microlensing. This distribution of microlensing errors characterizes an additional uncertainty on the time delay measurement, which is reported separately and then combined in quadrature with the model uncertainty. Including this measurement of uncertainty due to microlensing significantly improves the overall time delay uncertainty characterization, which is otherwise consistently underestimated when microlensing is ignored (Figure \ref{fig:microlensing_errors}). While including the uncertainty due to microlensing certainly improves the overall error characterization, its deviations from Gaussian at $|\Delta_{\rm{True}}-\Delta_{\rm{Measured}}|/\sigma\gtrsim1.0$ arise from extreme cases of microlensing. The agreement could likely be improved by increasing the number of draws from the microlensing GPR posterior, which was limited due to computation time to 100 per SN, or by combining the fitting methods described in sections \ref{sub:delays_separate} and \ref{sub:delays_colors} (see section \ref{sub:case_studies}).

\

\

\subsection{Parallel Analysis}
\label{sub:delays_separate}
If the separate light curves of a multiply-imaged SN are each reasonably well-sampled, then SNTD is capable of making time delay measurements by fitting each of the light curves in parallel. The parameters for a given model are separated into those that are inherent to the SN and therefore constant across each image (i.e. redshift, light curve shape parameters, host galaxy extinction, etc.), and those that are impacted by lensing and therefore will be unique to each image (i.e. amplitude, time of peak, etc.). These parameter sets are denoted $\bf\theta_{SN}$ and $\bf\theta_L$, respectively. 

We use the method of nested sampling to efficiently sample the full parameter space for $\bf\theta_{SN}$ and $\bf\theta_L$. Nested Sampling is effectively a Markov Chain Monte Carlo (MCMC) simulation that provides an estimate of the posterior probability distribution, but nested sampling additionally allows one to calculate the integral of the distribution and is more robust for finding global maxima \citep{Skilling:2004}. SNCosmo contains a modified nested sampling algorithm, which SNTD uses
to identify a best fit model for each separate image.
The default is to employ uniform prior distributions,but any form for the prior probability could be used. The algorithm takes the best fit models identified separately, obtains a joint posterior for parameters in $\bf\theta_{SN}$, and marginalizes over the parameters in $\bf\theta_L$. The result is a model for each image, where the parameters in $\bf\theta_{SN}$ are the same for each image and the parameters in $\bf\theta_L$ are likely different. Particularly when fitting light curves with little information, this gives some constraints on the $\bf\theta_{SN}$ parameters by using the fact that the underlying light curve for each image must be the same,  so any observed differences are only from extrinsic lensing effects like magnification, microlensing, and a relative time delay.

\begin{figure}[ht!]
\centering
\includegraphics[width=.45\textwidth]{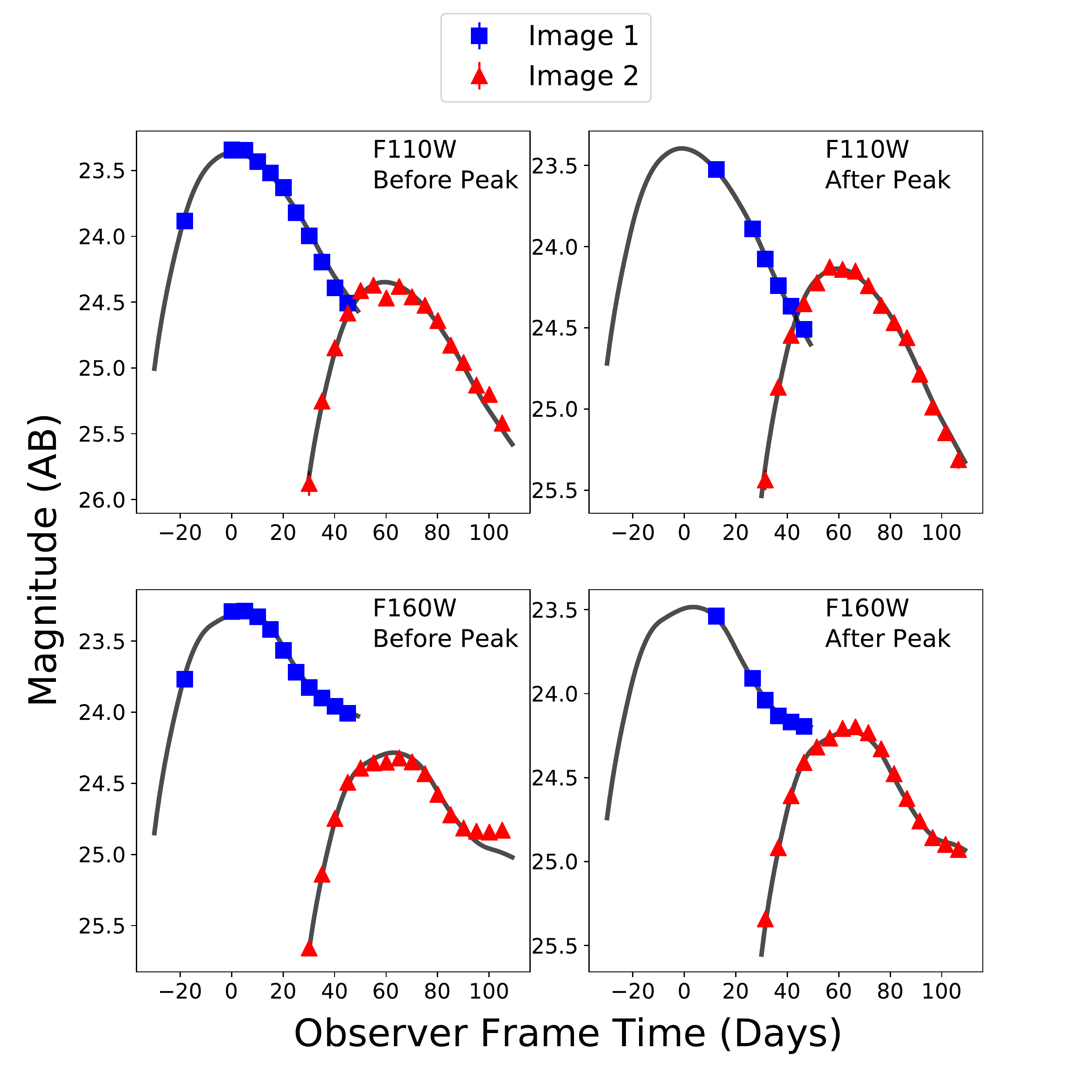}
\caption{\label{fig:delay_sep} An example of measuring the time delay for a simulated glSN Ia from section \ref{sub:sim_cases}, using the ``parallel'' fitting method outlined in this section. Each panel contains one of the images of this doubly-imaged SN, and the solid black line is the best-fit model. As is described in this section, the two light curves are fit in parallel, and then we obtain a joint posterior for parameters intrinsic to the SN light curve ($\theta_{SN}$) while marginalizing over extrinsic parameters ($\theta_L$).}
\end{figure}

Once the best-fit light curve parameters are identified, each image is fit once more with the $\bf\theta_{SN}$ parameters varying in a tightly bound range around the values measured above, while the $\bf\theta_L$ parameters remain unrestricted. This results in fine-tuned fits to each individual image (Figure \ref{fig:delay_sep}).

\subsection{Series Analysis}
\label{sub:delays_combined}
SNTD contains a second method for determining time delays that is particularly useful in the cases of relatively sparse light curves. This method differs from that of parallel fitting in section \ref{sub:delays_separate} in that the intrinsic light curve shape is identified first by fitting all or a subset of the light curves, which is then used to measure the time delays of the multiple images simultaneously. In both cases, we utilize the knowledge that the underlying light curve is the same for every image, but in this case we apply that leverage more directly by employing only a single set of intrinsic SN light curve parameters $\theta_{SN}$ (see section \ref{sub:delays_separate}) for the light curve model, in order to get simultaneous constraints on the relative time delays and magnifications for every image.

\begin{figure}[ht!]
\centering
\includegraphics[width=.45\textwidth]{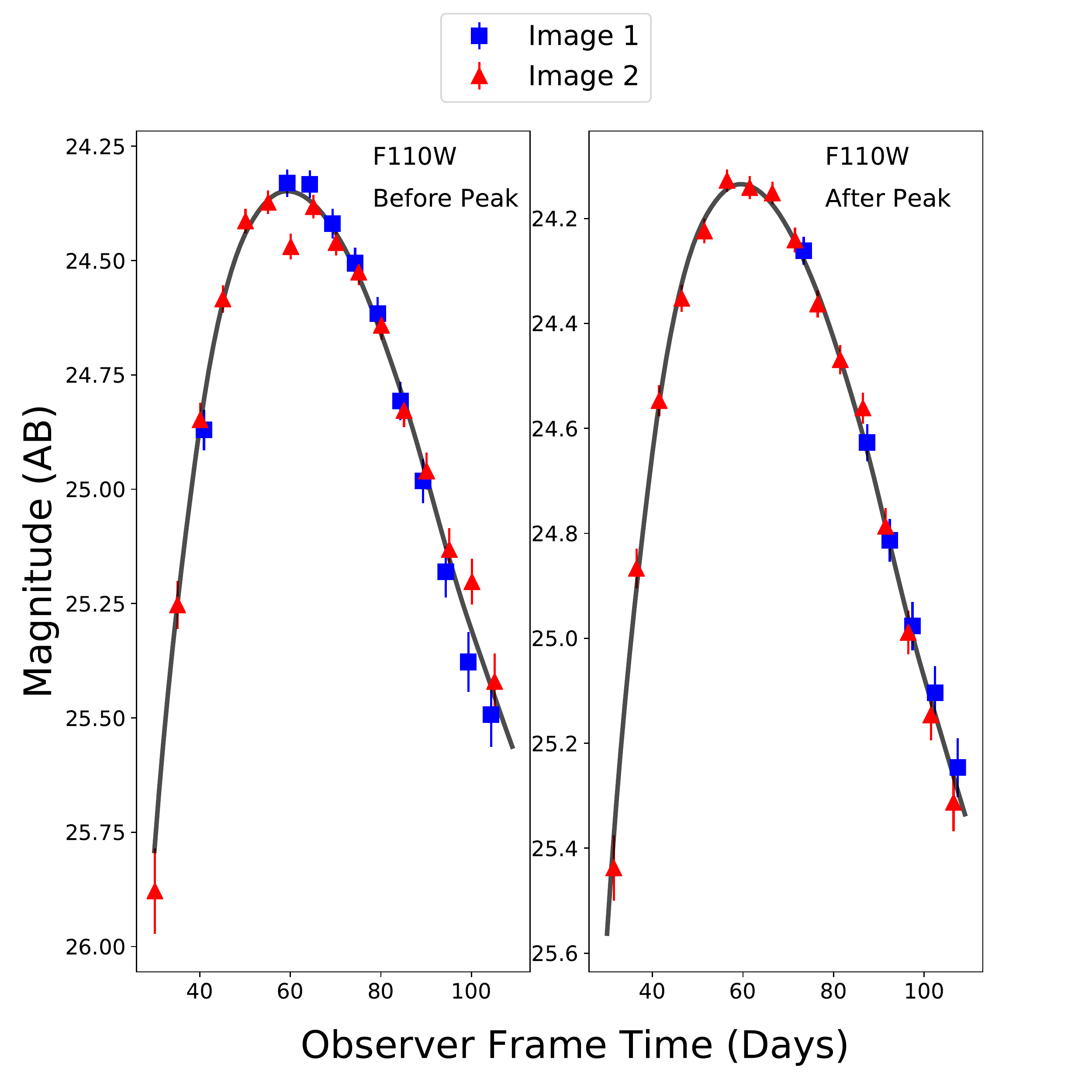}
\caption{\label{fig:delay_combined} An example of measuring the time delay for a simulated glSN Ia from section \ref{sub:sim_cases}, using the ``series'' measurement method outlined in this section. The left panel shows the simulated photometry for the before-peak case, and the right panel shows the simulated photometry for the after-peak case. In both panels, blue data points correspond to the leading image and red data points correspond to the trailing image. The black curves are the best-fit models, which are realizations of the intrinsic light curve shape. As is described in this section, the two light curves are fit together with this single model, making it a valuable method for somewhat sparsely sampled observed light curves.}
\end{figure}

To perform the series time delay measurement, SNTD effectively employs a double layered MCMC simulation that instead uses nested sampling to estimate the posterior probability distributions (see section \ref{sub:delays_separate}). The algorithm varies the $\bf\theta_{SN}$ parameters in the outer layer and the relative magnifications and time delays in the inner layer, attempting to simultaneously shift (time) and scale (flux) the photometric data from all images and fit it with a single varying light curve model (Figure \ref{fig:delay_combined}). As in the parallel approach, the default for this customized ``double-nested'' sampling is to use uniform priors for all parameters, though in principle any informative priors could be used.

\subsection{Color Curve Analysis}
\label{sub:delays_colors}
It has been proposed that the effect of microlensing on (at least) SNe Ia contains an achromatic period, meaning that it is not wavelength-dependent during this time \citep{Goldstein:2018}. Therefore color curves, which define the observed broad-band photometric color as a function of time, might be an effective way of measuring time delays and minimizing the complications of microlensing, as each filter should be affected identically. However, this method is only useful if the SN images are observed with multi-band photometry, the color curves are not featureless, and the achromatic phase is a true effect.

\begin{figure}[ht!]
\centering
\includegraphics[width=.45\textwidth]{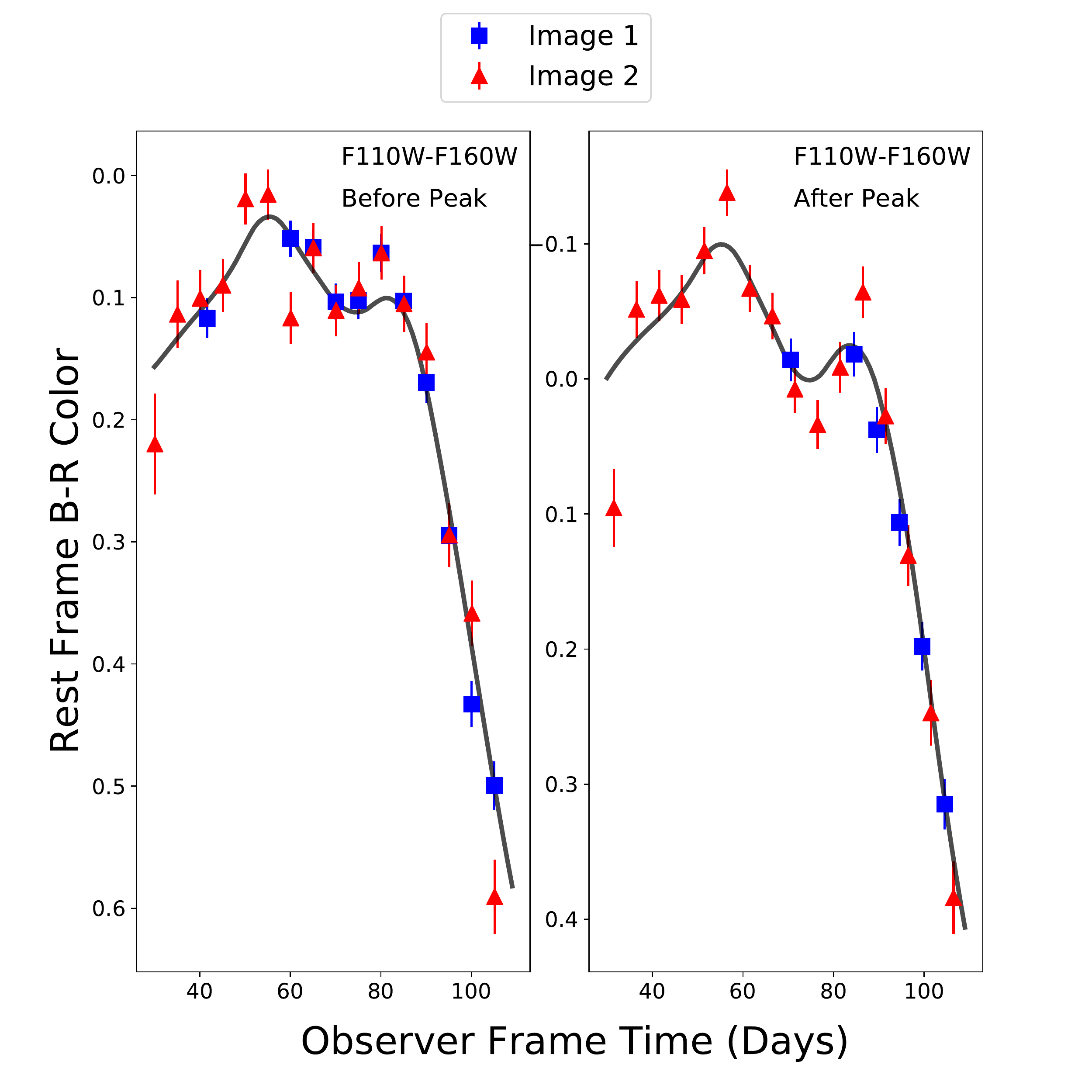}
\caption{\label{fig:delay_colors} An example of measuring the time delay for a simulated glSN Ia from section \ref{sub:sim_cases}, using the ``color curve'' method outlined in this section. The left panel shows the simulated photometry for the before-peak case, and the right panel shows the simulated photometry for the after-peak case. In both panels, the blue squares correspond to the leading image and the red triangles to the trailing image. The solid black curves are the best-fit color curve model, and the epoch of peak B-band brightness is at 60 days.}
\end{figure}

In cases where the SN classification is well-known, existing templates can be used to obtain expected color curves, which can then be used to constrain the functions being used to fit the data (Figure \ref{fig:delay_colors}). Otherwise, we must resort to flexible functions such as splines or Chebyshev polynomials to fit the data. In either case, we can still leverage the fact that the color curve for each image of the SN should be the same, as in section \ref{sub:delays_combined}. Therefore, the same ``double-nested'' sampling algorithm is employed to obtain time delays from color curves (see section \ref{sub:delays_combined}). The algorithm once again attempts to shift (time) and scale (flux) the color curve data for all the SN images, while fitting the combined data with a single varying color curve model. SNTD allows the color curve flux data to be individually scaled in order to account for chromatic parameters in $\bf\theta_L$ (see section \ref{sub:delays_separate}) that will affect each image differently, such as dust extinction or non-achromatic microlensing.

\

\section{SNTD: Case Studies}
\label{sec:cases}
\label{sub:case_studies}
\begin{figure}[h!]
\centering
\includegraphics[width=.45\textwidth]{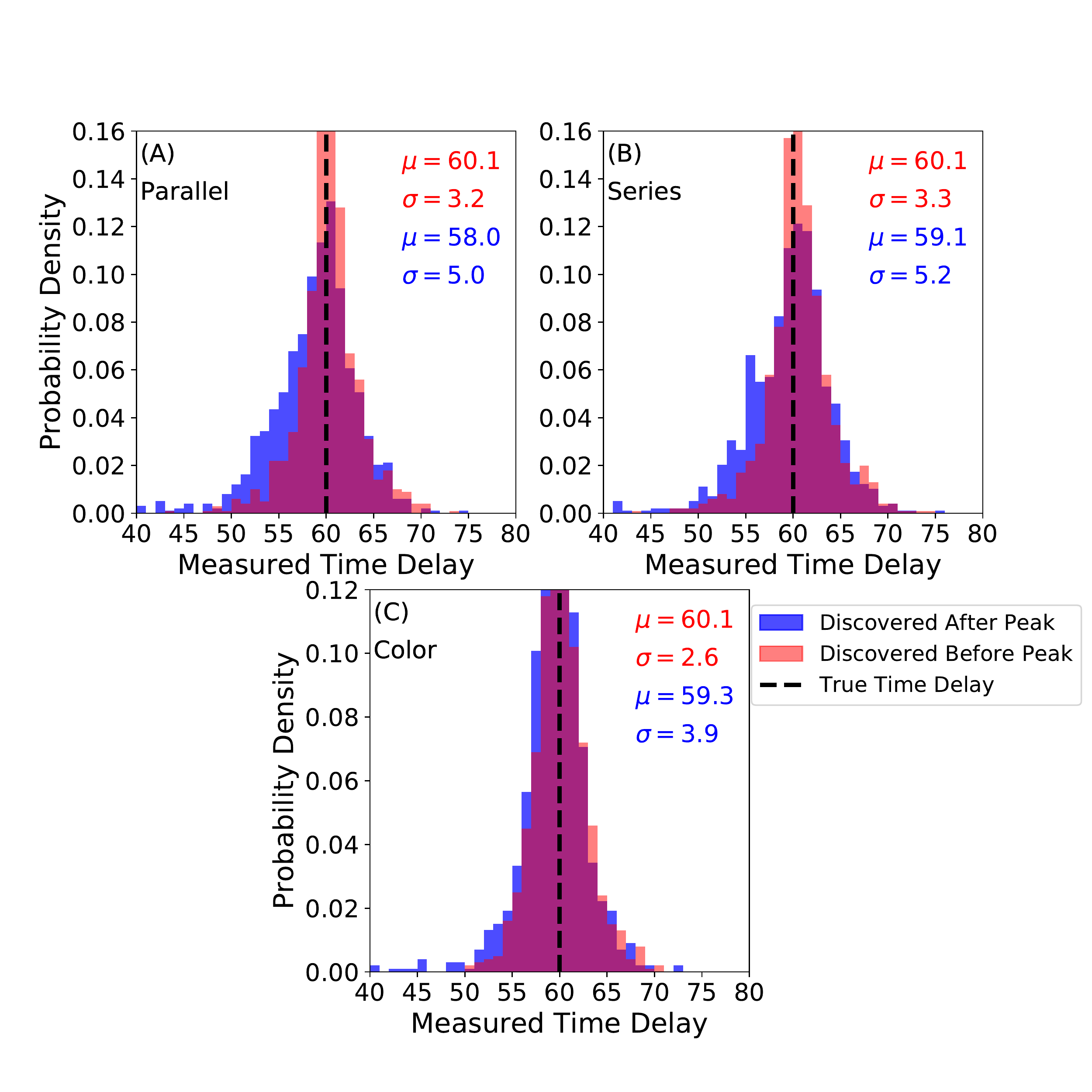}
\caption{\label{fig:case_results} In each panel, red corresponds to the before-peak case and blue to the after-peak case. A black dashed line marks the true time delay for each study. The mean ($\mu$) and standard deviation ($\sigma$) of each distribution is shown in the upper right of each panel in corresponding colors. Panels A, B, and C show results for the parallel, series, and color curve routines respectively, all described in section \ref{sec:delays}.}
\end{figure}

We measure the relative time delays of the 2,000 simulated glSNe described in section \ref{sub:sim_cases}, exploiting our knowledge of the SN classification by using SED templates to perform the light curve fitting. The simulations are separated into SNe ``discovered'' before peak and after peak brightness. We obtained three different sets of results for each case by using the methods described in sections \ref{sub:delays_separate}-\ref{sub:delays_colors} to measure the time delays, which are summarized in table \ref{tab:case_results} and figure \ref{fig:case_results}. All three fitting methods measured with nearly identical accuracy and for the before-peak case, which is to be expected for these simulations as the light curves were sufficiently well-sampled to use any approach. Still, the fact that achromatic microlensing (used for these simulations, see section \ref{sub:micro}) does not affect the color curve method led to a higher precision. The series fitting routine did perform slightly better in the after-peak scenario when compared to the parallel routine, reflecting that it can be more robust in cases where one or more images has a lower quality light curve. Nevertheless, the accuracy decreased in the after-peak case study for both of these techniques, the systematic offset stemming from the lack of an inflection point to identify peak brightness, leading to an underestimation of the time delay. Likewise the precision for each technique is reduced for the after-peak case. 

The difference between the two cases for the color curve routine is much less pronounced, corresponding to the relative independence of a color curve on peak brightness. While the lack of before-peak data does not directly harm the color curve fitting, it still leads to a decrease in accuracy as most of the simulated leading-image light curves only contained 3 or 4 data points before falling below the detection threshold (Figure \ref{fig:case_sims}). The performance of the color curve fitting method is somewhat idealized here however, as the simulated microlensing is achromatic and therefore completely absent in each color curve. It's worth noting that the color curve routine's relative insensitivity to peak brightness and comparatively small number of catastrophic outliers may suggest that combining the color curve fitting routine with either the parallel or series fitting methods could enable more accurate measurements. Such a reduction in the tails of figure \ref{fig:case_results}A-B should also lead to an improvement in the uncertainty characterization described in section \ref{sub:micro_fitting} and figure \ref{fig:microlensing_errors}.

\renewcommand{\arraystretch}{1.1}

\begin{table}[h!]
\centering
\begin{tabular}{c|crr}
\toprule
\multicolumn{1}{c}{\textbf{Simulation}}  &\multicolumn{1}{c}{\textbf{Method}}  & \multicolumn{1}{c}{$\mu_{\Delta t}$} &\multicolumn{1}{c}{$\sigma_{\Delta t}$}  \\
\hline
\multirow{3}{*}{Before Peak}&Parallel (Section \ref{sub:delays_separate})&60.1&3.2\\
&Series (Section \ref{sub:delays_combined})&60.1&3.3\\
&Color Curves (Section \ref{sub:delays_colors})&60.1&2.6\\
\hline
\multirow{3}{*}{After Peak}&Parallel (Section \ref{sub:delays_separate})&58.0&5.0\\
&Series (Section \ref{sub:delays_combined})&59.1&5.2\\
&Color Curves (Section \ref{sub:delays_colors})&59.3&3.9\\\hline

\end{tabular}
\caption{\label{tab:case_results} The results of each time delay measurement technique for the before-peak and after-peak case studies. As expected, there is a decrease in accuracy and precision when you lose before-peak observations due to a lack of single inflection point and a decrease in the number of light curve points to fit.}
\end{table}

\pagebreak

\section{Discussion and Future Work}
\label{sec:discussion}
We present an open source software package, SNTD, specifically optimized for the analysis of glSNe. The package has both simulation and time delay measurement capabilities, and has a range of methods available to perform these tasks. In this paper, we use the simulation component of SNTD to create two case studies of a targeted survey galaxy, similar to the observation strategy of \citet{Shu2018a}. In the first case study the leading image of each glSN is always discovered before peak, while in the second case study the leading image of each glSN only contains after-peak observations.  Using these simulated SNe, we present the time delay measurement capabilities of SNTD by running its automated fitting algorithm on the simulated data using a variety of tools present in the SNTD package, and reporting the respective measurement accuracy and precision. We find that each routine has different strengths and weaknesses, but that overall obtaining before-peak observations increases the accuracy of time delay measurements by $\sim1$ day and the precision by $\sim2$ days ($2\%$ and $3\%$ for this work, respectively). While these effects seem small, they would be considerable for the many time delays expected from future surveys that will be on the order of hours or days \citep{Goldstein:2018b,Huber:2019}. 

The SNTD package is being applied to the case of glSN Refsdal in a parallel paper, to improve upon the measured time delays presented by \citet{Rodney:2016} and \citet{Kelly:2016}. SNTD is designed to be useful for time delay measurements of the large sample of glSNe expected from LSST and WFIRST, and can currently be used to simulate the constraints on $H_0$ possible from these future surveys. As we move into the next decade the number of glSN discoveries will increase by orders of magnitude, and SNTD provides valuable capabilities for maximizing the impact of this new era of SN time delay cosmography.

\

\

\

ACKNOWLEDGEMENTS:
 
 \ 
 
 \noindent The authors thank Vivien Bonvin, Patrick Kelly, and Tommaso Treu for very helpful discussion of this paper. We would also like to thank the referee for their insightful comments, which certainly strengthened this work.
 
 Support for this work was provided by NASA through grant number HST-AR-15050 from the Space Telescope Science Institute, which is operated by AURA, Inc., under NASA contract NAS 5-26555.

 \pagebreak

\bibliographystyle{aas}

\pagebreak

\appendix
\setcounter{table}{0}
\renewcommand{\thetable}{A\arabic{table}}

\setcounter{figure}{0}
\renewcommand{\thefigure}{A\arabic{figure}}

\begin{longtable}{ccc} 
\textbf{Model Name} & \textbf{Type} & \textbf{Reference} \\
\hline
nugent-sn1a	&	SN Ia	&	\citet{Nugent:2002}	\\
nugent-sn91t	&	SN Ia	&	\citet{Stern:2004}	\\
nugent-sn91bg	&	SN Ia	&	\citet{Nugent:2002}	\\
nugent-sn1bc	&	SN Ib/c	&	\citet{Levan:2005}	\\
nugent-hyper	&	SN Ib/c	&	\citet{Levan:2005}	\\
nugent-sn2p	&	SN IIP	&	\citet{Gilliland:1999}	\\
nugent-sn2l	&	SN IIL	&	\citet{Gilliland:1999}	\\
nugent-sn2n	&	SN IIn	&	\citet{Gilliland:1999}	\\
s11-2004hx	&	SN IIL/P	&	\citet{Sako:2011}	\\
s11-2005lc	&	SN IIP	&	\citet{Sako:2011}	\\
s11-2005hl	&	SN Ib	&	\citet{Sako:2011}	\\
s11-2005hm	&	SN Ib	&	\citet{Sako:2011}	\\
s11-2005gi	&	SN IIP	&	\citet{Sako:2011}	\\
s11-2006fo	&	SN Ic	&	\citet{Sako:2011}	\\
s11-2006jo	&	SN Ib	&	\citet{Sako:2011}	\\
s11-2006jl	&	SN IIP	&	\citet{Sako:2011}	\\
hsiao	&	SN Ia	&	\citet{Hsiao:2007}	\\
hsiao-subsampled	&	SN Ia	&	\citet{Hsiao:2007}	\\
salt2	&	SN Ia	&	\citet{Guy:2010}	\\
salt2	&	SN Ia	&	\citet{Guy:2007}	\\
salt2	&	SN Ia	&	\citet{Betoule:2014}	\\
salt2-extended	&	SN Ia	& \citet{Pierel2018}		\\
snf-2011fe	&	SN Ia	&	\citet{Pereira:2013}	\\
snana-2004fe	&	SN Ic	&	\citet{Kessler:2010}\\
snana-2004gq	&	SN Ic	&	\citet{Kessler:2010}	\\
snana-sdss004012	&	SN Ic	&	\citet{Kessler:2010}	\\
snana-2006fo	&	SN Ic	&	\citet{Kessler:2010}	\\
snana-sdss014475	&	SN Ic	&	\citet{Kessler:2010}	\\
snana-2006lc	&	SN Ic	&	\citet{Kessler:2010}	\\
snana-2007ms	&	SN II-pec	&		\citet{Kessler:2010}\\
snana-04d1la	&	SN Ic	&	\citet{Kessler:2010}	\\
snana-04d4jv	&	SN Ic	&	\citet{Kessler:2010}	\\
snana-2004gv	&	SN Ib	&	\citet{Kessler:2010}	\\
snana-2006ep	&	SN Ib	&\citet{Kessler:2010}		\\
snana-2007y	&	SN Ib	&	\citet{Kessler:2010}	\\
snana-2004ib	&	SN Ib	&	\citet{Kessler:2010}	\\
snana-2005hm	&	SN Ib	&	\citet{Kessler:2010}	\\
snana-2006jo	&	SN Ib	&\citet{Kessler:2010}		\\
snana-2007nc	&	SN Ib	&	\citet{Kessler:2010}	\\
snana-2004hx	&	SN IIP	&	\citet{Kessler:2010}	\\
snana-2005gi	&	SN IIP	&	\citet{Kessler:2010}	\\
snana-2006gq	&	SN IIP	&\citet{Kessler:2010}		\\
snana-2006kn	&	SN IIP	&\citet{Kessler:2010}		\\
snana-2006jl	&	SN IIP	&\citet{Kessler:2010}		\\
snana-2006iw	&	SN IIP	&	\citet{Kessler:2010}	\\
snana-2006kv	&	SN IIP	&	\citet{Kessler:2010}	\\
snana-2006ns	&	SN IIP	&\citet{Kessler:2010}		\\
snana-2007iz	&	SN IIP	&	\citet{Kessler:2010}	\\
snana-2007nr	&	SN IIP	&\citet{Kessler:2010}		\\
snana-2007kw	&	SN IIP	&\citet{Kessler:2010}		\\
snana-2007ky	&	SN IIP	&\citet{Kessler:2010}		\\
snana-2007lj	&	SN IIP	&\citet{Kessler:2010}		\\
snana-2007lb	&	SN IIP	&\citet{Kessler:2010}		\\
snana-2007ll	&	SN IIP	&\citet{Kessler:2010}		\\
snana-2007nw	&	SN IIP	&\citet{Kessler:2010}		\\
snana-2007ld	&	SN IIP	&	\citet{Kessler:2010}	\\
snana-2007md	&	SN IIP	&\citet{Kessler:2010}		\\
snana-2007lz	&	SN IIP	&	\citet{Kessler:2010}	\\
snana-2007lx	&	SN IIP	&	\citet{Kessler:2010}	\\
snana-2007og	&	SN IIP	&	\citet{Kessler:2010}	\\
snana-2007ny	&	SN IIP	&	\citet{Kessler:2010}	\\
snana-2007nv	&	SN IIP	&	\citet{Kessler:2010}	\\
snana-2007pg	&	SN IIP	&	\citet{Kessler:2010}	\\
snana-2006ez	&	SN IIn	&	\citet{Kessler:2010}	\\
snana-2006ix	&	SN IIn	&	\citet{Kessler:2010}	\\
whalen-z15b	&	PopIII	&\citet{Whalen:2013}	\\
whalen-z15d	&	PopIII	&	\citet{Whalen:2013}	\\
whalen-z15g	&	PopIII	&\citet{Whalen:2013}	\\
whalen-z25b	&	PopIII	&\citet{Whalen:2013}	\\
whalen-z25d	&	PopIII	&\citet{Whalen:2013}	\\
whalen-z25g	&	PopIII	&\citet{Whalen:2013}	\\
whalen-z40b	&	PopIII	&\citet{Whalen:2013}	\\
whalen-z40g	&	PopIII	&	\citet{Whalen:2013}	\\
mlcs2k2	&	SN Ia	&	\citet{Jha:2007}	\\
\caption{The SED models present in the SNCosmo package to describe SN evolution with wavelength and time.}
\label{Atab:sncosmo_seds}
\end{longtable}

\begin{longtable}{crrrrc} 
\textbf{Band} & \textbf{Time} & \textbf{Gain} & \textbf{Sky Noise }& \textbf{ZP} & \textbf{ZP System} \\
 & (Days) &  & (Counts) &  &  \\\hline
F125W & 0.0 & 50.0 & 0.008 & 26.3 & AB \\
F125W & 3.0 & 50.0 & 0.028 & 26.3 & AB \\
F125W & 6.0 & 50.0 & 0.008 & 26.3 & AB \\
F125W & 9.0 & 50.0 & 0.026 & 26.3 & AB \\
F125W & 12.0 & 50.0 & 0.011 & 26.3 & AB \\
F125W & 15.0 & 50.0 & 0.026 & 26.3 & AB \\
F125W & 18.0 & 50.0 & 0.013 & 26.3 & AB \\
F125W & 21.0 & 50.0 & 0.014 & 26.3 & AB \\
F125W & 24.0 & 50.0 & 0.029 & 26.3 & AB \\
F125W & 27.0 & 50.0 & 0.007 & 26.3 & AB \\
F125W & 30.0 & 50.0 & 0.029 & 26.3 & AB \\
F125W & 33.0 & 50.0 & 0.008 & 26.3 & AB \\
F125W & 36.0 & 50.0 & 0.024 & 26.3 & AB \\
F125W & 39.0 & 50.0 & 0.025 & 26.3 & AB \\
F125W & 42.0 & 50.0 & 0.021 & 26.3 & AB \\
F125W & 45.0 & 50.0 & 0.008 & 26.3 & AB \\
F125W & 48.0 & 50.0 & 0.010 & 26.3 & AB \\
F125W & 51.0 & 50.0 & 0.006 & 26.3 & AB \\
F125W & 54.0 & 50.0 & 0.013 & 26.3 & AB \\
F125W & 57.0 & 50.0 & 0.025 & 26.3 & AB \\
F125W & 60.0 & 50.0 & 0.012 & 26.3 & AB \\
F125W & 63.0 & 50.0 & 0.024 & 26.3 & AB \\
F125W & 66.0 & 50.0 & 0.007 & 26.3 & AB \\
F125W & 69.0 & 50.0 & 0.019 & 26.3 & AB \\
F125W & 72.0 & 50.0 & 0.025 & 26.3 & AB \\
F125W & 75.0 & 50.0 & 0.024 & 26.3 & AB \\
F160W & 0.0 & 50.0 & 0.009 & 26.0 & AB \\
F160W & 3.0 & 50.0 & 0.028 & 26.0 & AB \\
F160W & 6.0 & 50.0 & 0.027 & 26.0 & AB \\
F160W & 9.0 & 50.0 & 0.009 & 26.0 & AB \\
F160W & 12.0 & 50.0 & 0.008 & 26.0 & AB \\
F160W & 15.0 & 50.0 & 0.011 & 26.0 & AB \\
F160W & 18.0 & 50.0 & 0.025 & 26.0 & AB \\
F160W & 21.0 & 50.0 & 0.014 & 26.0 & AB \\
F160W & 24.0 & 50.0 & 0.027 & 26.0 & AB \\
F160W & 27.0 & 50.0 & 0.024 & 26.0 & AB \\
F160W & 30.0 & 50.0 & 0.025 & 26.0 & AB \\
F160W & 33.0 & 50.0 & 0.019 & 26.0 & AB \\
F160W & 36.0 & 50.0 & 0.015 & 26.0 & AB \\
F160W & 39.0 & 50.0 & 0.026 & 26.0 & AB \\
F160W & 42.0 & 50.0 & 0.022 & 26.0 & AB \\
F160W & 45.0 & 50.0 & 0.026 & 26.0 & AB \\
F160W & 48.0 & 50.0 & 0.029 & 26.0 & AB \\
F160W & 51.0 & 50.0 & 0.025& 26.0 & AB \\
F160W & 54.0 & 50.0 & 0.017 & 26.0 & AB \\
F160W & 57.0 & 50.0 & 0.008 & 26.0 & AB \\
F160W & 60.0 & 50.0 & 0.028 & 26.0 & AB \\
F160W & 63.0 & 50.0 & 0.010 & 26.0 & AB \\
F160W & 66.0 & 50.0 & 0.006 & 26.0 & AB \\
F160W & 69.0 & 50.0 & 0.016 & 26.0 & AB \\
F160W & 72.0 & 50.0 & 0.027 & 26.0 & AB \\
F160W & 75.0 & 50.0 & 0.019 & 26.0 & AB \\
\caption{An example of the observation tables used to produce simulated light curves section \ref{sub:sim_cases}. Each SN is observed on a timescale of 0-100 Days, with a cadence of 3 days \citep{Shu2018a} for a total of 25 epochs.}
\label{Atab:observations}
\end{longtable}

\

\begin{table}[!ht]
\centering
\begin{tabular}{cccc}
\textbf{Type} & \textbf{$M_R$} & \textbf{$\sigma$} & \textbf{Source} \\
\hline
Ia&-19.37&0.47&\citep{WangL:2006} \\
Ib&-17.90&0.90&\citep{Drout:2011}\\
Ic&-18.30&0.60&\citep{Drout:2011}\\
IcBL&-19.00&1.10&\citep{Drout:2011}\\
II-P&-16.56&0.80&\citep{Li:2011a}\\
II-L&-17.66&0.42&\citep{Li:2011a}\\
IIn&-18.25&1.00&\citep{Kiewe:2012}\\
\end{tabular}
\caption{The Gaussian luminosity functions for each SN type, presented in table 3 of \citet{Graur:2014}.}
\label{Atab:graur_luminosity}
\end{table}

\

\begin{table}[!ht]
\centering
\begin{tabular}{ccc}
\textbf{Type} & \textbf{$M_B$} & \textbf{$\sigma$} \\
\hline
Ia&-19.25$\pm$0.20&0.50\\
Ib&-17.45$\pm$0.33&1.12\\
Ic&-17.66$\pm$0.40&1.18\\
IIb&-16.99$\pm$0.45&0.92\\
II-L&-17.98$\pm$0.34&0.86\\
II-P&-16.75$\pm$0.37&0.98\\
IIn&-18.53$\pm$0.32&1.36\\
\end{tabular}
\caption{The Gaussian luminosity functions for each SN type, adapted from table  of \citet{Richardson:2014}.}
\label{Atab:richardson_luminosity}
\end{table}

\end{document}